\documentclass[letterpaper]{article}
\usepackage{graphicx}
\usepackage{cite}
\graphicspath{ {LaTeX/} }
\begin{document}
\textbf{\huge\\Electroweak phase transition and some related phenomena- a brief review}\\
\\Buddhadeb Ghosh \\Centre for Advanced Studies, Department of Physics, The University of  Burdwan, Burdwan – 713 104.\\
\\Email: bghosh@phys.buruniv.ac.in        ghoshphysics@yahoo.co.in
\begin{abstract}
In this article, we give a bird’s eye view of research on electroweak phase transition and some related phenomena, viz., cosmological baryogenesis, electroweak bubble dynamics and generation of gravitational waves. Our presentation revolves around the observation that a strong first order electroweak phase transition cannot be obtained in the Standard Model for experimentally-favoured Higgs mass  and hence the cosmological events associated with this kind of phase transition cannot be explained in this model.  However, this phase transition can be achieved in a number of Beyond Standard Models.  As a prototype case , we consider the littlest Higgs model with T parity and show the results of some calculations within this model.\par
\end{abstract}
\section{Introduction}
According to the Standard Big Bang Cosmology, the Universe underwent an electroweak phase transition (EWPT), associated with  the  spontaneous symmetry breaking (SSB): $SU(2)_L\otimes U(1)_Y\rightarrow U(1)_{EM}$ at $t\sim10^{-11} s$ and $T\sim300 GeV$\cite{1}. The weak gauge bosons and fermions obtained masses by the electroweak SSB and the Higgs mechanism, whereas the EWPT is believed to be connected with the cosmological processes, such as the mechanism of matter-antimatter asymmetry generation i.e., baryogenesis, production of gravitational waves, formation of topological defects like the cosmic strings or the domain walls etc.\par
\setlength{\parindent}{.5 in}Although the Standard Model (SM) of particle physics was quite efficient in predicting masses of gauge bosons and generating the masses of fermions by SSB and many other experimental observations at the 100 GeV scale, it, as against  the Beyond Standard Models (BSMs) such as supersymmetric, extra-dimensional, two Higgs doublet, little Higgs models etc.,  proved to be insufficient in quantitative assessment of  the above-mentioned cosmological events as well as accounting for the existence of dark matter and dark energy. However, the mass of the recently discovered \cite{2}  Higgs-like particle at the Large Hadron Collider (LHC), viz., 126 GeV, is not far off from the lower bound of the Higgs boson mass, 115 GeV, obtained in the LEP experiments\cite{3}.  This observation has fuelled the speculation that even the scalar sector of the electroweak theory can be described satisfactorily by an appropriate minimal extension of SM. This is testified by some recent works to study the EWPT within some extensions of the SM, both in the pre-\cite{4} and post-\cite{5} Higgs discovery period.\par
\setlength{\parindent}{.5 in}In this article, we briefly present an update of  the research and evolution of ideas in EWPT, baryogenesis, bubble dynamics and generation of gravitational waves at the electroweak scale in the SM and some BSMs. The style of our presentation is to convey important results qualitatively. Details of models and their calculational frameworks can be obtained from the references which have been cited here.\par
\setlength{\parindent}{.5 in}   In section 2, we present a picture of the development of knowledge of electroweak phase transition starting from the early days of research in this field. In section 3, we discuss baryogenesis, stressing the point that a strong first-order electroweak phase transition is necessary for baryogenesis. We also present a new baryogenesis scenario obtained within the littlest Higgs model with T parity. In section 4, we mention some aspects of the generation of gravitational waves in the early Universe. Finally, in section 5, we write some concluding remarks.
\section{The Electroweak Phase Transition}
Theoretical studies related to the EWPT have appeared in the literature for the last four decades or so. Among the early works, we may mention, the phase transition in finite-temperature gauge theories\cite{6}, SSB in massless finite-temperature field theories\cite{7}, dependence of the behavior of the cosmological phase transition on the Higgs mass $m_H$\cite{8}, radiative effects on SSB\cite{9}, cosmological consequence of  a Coleman-Weinberg type EWPT \cite{10} and its impact on the expansion of the Universe\cite{11}.\par
\setlength{\parindent}{.5 in}The usual framework for studying the EWPT is the finite-temperature effective potential (FTEP) of the Higgs field. An effective potential (EP) \cite{12}, which can be calculated at various loop orders, can be thought to be a quantum-corrected classical potential. It may be mentioned that the EP which we consider here is gauge-dependent. It is calculated in the Landau gauge\cite{12} which has the merit of decoupling the unphysical degrees of freedom from the theory. Figure 1 shows the diagrammatic representation of the one-loop order EP. The EP which takes care of quantum fluctuations in the potential of a scalar field has been found to be quite useful in the quantum theory of SSB\cite{9} as the fluctuations may render the mass squared parameter in the classical potential negative, causing the SSB. It is interesting to note that the EP generated by quantum fluctuations has the interpretation\cite{13} of the fluctuating energy density, similar to the zero point energy of the harmonic oscillator. \par
\setlength{\parindent}{.5 in}    By studying the effective potential and the associated thermodynamic quantities in a field-theoretic model, one can ascertain whether there is a phase transition and what is the order of the phase transition. The behaviors of FTEP near the transition temperature $T_c$ in the case of first and second-order phase transitions are shown schematically in Figure 2. It is crucial to determine the order of EWPT, because the early Universe phenomena at the electroweak scale will depend on this order if there is an EWPT.\par
\setlength{\parindent}{.5 in}In the SM, it was shown \cite{14} that there is substantial $(\sim20-40\%)$ two-loop correction to the one-loop FTEP, which therefore questioned the validity of perturbative method in the evaluation of FTEP. To take into account the
\begin{figure}
  \centering
  \includegraphics{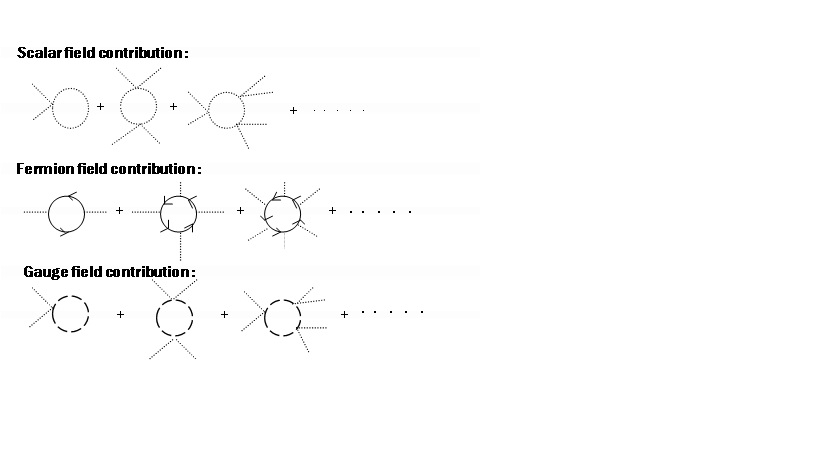}
  \caption{Diagrammatic representations of one-loop order effective potential, which is quantum correction to classical potential.}\label{Figure 1}
\end{figure}
nonperturbative effects, many lattice calculations \cite{15,16,17,18,19,20,21} were done which yielded more reliable results than with the perturbative loop calculations.\par
\setlength{\parindent}{.5 in}The main result of the lattice calculations was that for weakly-coupled electroweak theory there is first-order EWPT up to $m_H\cong 80 GeV$,  but there is no first or second-order EWPT beyond this value,  rather there is a sharp cross-over characterized by a rapid increase of the order parameter, $<\phi^\dagger\phi>$ below the transition temperature \cite{19}. The results of lattice numerical simulations were in agreement with thermodynamic calculations \cite{21}.
\begin{figure}[h]
  \centering
  \includegraphics{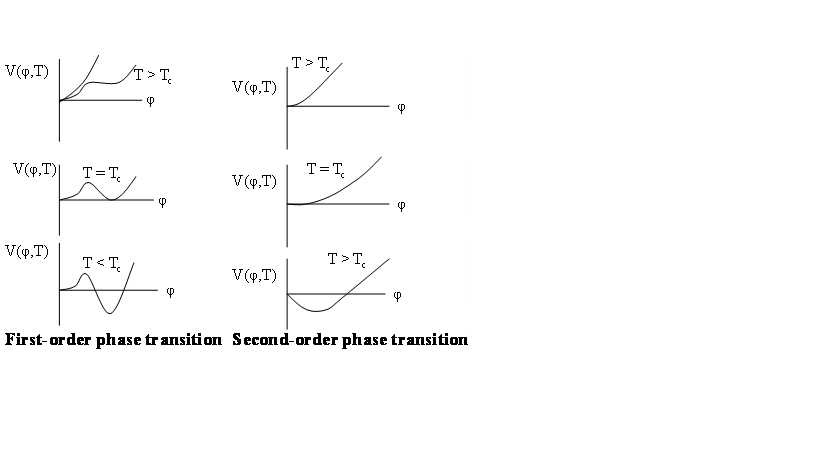}
  \caption{Variations of the effective potential with temperature in the case of first and second order phase transitions}\label{Figure 2}
\end{figure}
It may be mentioned here that some authors, while presenting a baryogenesis scenario, have considered second-order EWPT associated either with evaporation of primary black holes \cite{22} or with dynamics of cosmic strings \cite{23}. However, as was argued in Ref.\cite{24}, a second-order EWPT in general cannot produce the observed baryon to entropy ratio, as it cannot create the necessary thermal nonequilibrium condition.\par
\setlength{\parindent}{.5 in}  Coming back to the case of EWPT in SM, it is established that a strong first-order electroweak phase transition (SFOEWPT), as required by successful baryogenesis in the early Universe (to be discussed in detail in the next section), is not possible in this model. This fact led researchers to explore the possibility of SFOEWPT in various BSMs such as supersymmetric, two Higgs doublet, extra-dimensional and little Higgs models.\par
\setlength{\parindent}{.5 in} Within the framework of minimally supersymmetric standard model (MSSM) \cite{25}, SFOEWPT has been reported to be found in studies involving FTEP at one or two loop orders \cite{26}, in nonperturbative lattice simulations \cite{27}, with a fourth generation of particles \cite{28}, in an U(1)-extended version \cite{29} as well as in an R-parity violating \cite{25} extension \cite{30}. Recently, possibility of SFOEWPT has been examined \cite{31} in the next-to-minimal supersymmetric standard model (NMSSM) \cite{25} and in models \cite{32} involving dark matter candidates and also in two-Higgs doublet model \cite{33} and extra-dimensional models\cite{34,35}.\par
\setlength{\parindent}{.5 in} One of the intriguing phenomena in the context of first-order electroweak phase transition is the existence of broken phase at high temperature and inverse symmetry breaking, which have been seen in an extra-dimensional model\cite{35} and in a finite-temperature version \cite{36} of the littlest Higgs model with $T$ parity(LHT)\cite{37}. In the LHT model, the Higgs fields are subsets of pseudo Nambu Goldstone bosons contained in a nonlinear sigma field, $\Sigma$. The periodic structure of $\Sigma$ gives a global structure of FTEP as a function of the physical Higgs field, $h$. The variation of the FTEP with temperature is shown in Fig.3. We observe an inverse SFOEWPT with $h(T_c)/T_c\cong1.2$. The significance of this quantity, in the context of baryogenesis, will be discussed in the next section. In the parameter space, where the SFOEWPT is observed, $m_H\cong156  GeV$.\par
\setlength{\parindent}{.5 in}In view of the recent LHC observation of $m_H\cong126  GeV$, researchers, in coming days, may still focus on extensions of SM {\cite{4,5}, with possible mechanisms for stabilizing the Higgs vacuum.\par
\setlength{\parindent}{.5 in} In the context of dynamics of EWPT, a difference of between the SM and a broad class of BSMs including the LHT can be understood from a recent study \cite {38}, where it has been shown that there is a strong correlation between this dynamics and the cubic Higgs self-coupling. The BSMs showing an SFOEWPT would predict a large deviation of this coupling from the SM-predicted value.
\section{Baryogenesis}
The motivation behind the search for the SFOEWPT in a field theoretic model is to simulate an early universe scenario with a thermal non-equilibrium situation at the time of electroweak phase transition which was instrumental in preventing a washout of a generated baryon-antibaryon asymmetry, thus conforming to the observed baryon-antibaryon asymmetry \cite{39} in the present Universe.\par
\setlength{\parindent}{.5 in}There are several models of  baryogenesis (as many as forty four)\cite{40}, which we may broadly classify as, Planck- or string-scale baryogenesis \cite{41}, GUT-scale baryogenesis \cite{42} and electroweak baryogenesis \cite{43,44,45}, Of  these three classes, only the electroweak baryogenesis (EWBG) models can, understandably,  be tested by the present-day accelerator experiments.\par
\setlength{\parindent}{.5 in}Usually, any baryogenesis model has to satisfy Sakharov’s three conditions \cite{46}: (i) Baryon number violation, (ii) C and CP violation, (iii) Departure from thermal equilibrium. In the SM, large baryon number violation \cite{47} is possible at high
\begin{figure}[h]
  \centering
  \includegraphics{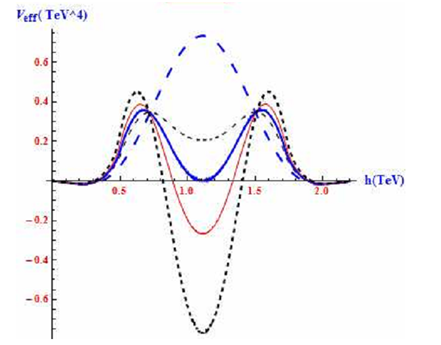}
  \caption{Finite-temperature effective potential in LHT model at temperatures (from top to bottom): $T=0 TeV$(blue, large-dashed), $T=0.85 TeV$ (black, medium-dashed), $T=0.925 TeV$ (blue, thick solid), $T=1 TeV$ (red solid) and $T=1.1 TeV$ (black, thick-dashed). The transition temperature is $T_c=0.925 TeV$. (Ref.\cite{36})(Reprinted with permission from [ S. Aziz, B.Ghosh and G. Dey, \textit{Physical Review} \textbf{D79}, 075001 (2009)] Copyright (2009) by the American Physical Society).}\label{Figure 3}
\end{figure}
temperature by sphaleron transitions \cite{48} between the degenerate vacua of the $SU(2)$ gauge field. Sphalerons \cite{48} are static but unstable solutions of classical field equations in the electroweak theory. However it is difficult to satisfy the second and the third conditions in the SM, because the CP violation in this model is too low to give the cosmological baryon to entropy ratio \cite{49} and this model does not show an SFOEWPT, as discussed in the previous section, for $m_H<32 GeV$ \cite{50}. These problems have been addressed and solutions  sought in some of the BSMs, such as the MSSM \cite{43} and its extensions\cite{44}, and the THDM\cite{45}.\par
\setlength{\parindent}{.5 in}In the context of electroweak phase transition and baryogenesis in the early Universe, an useful quantity to calculate is the ratio R=(baryon number violation  rate)/(Universe expansion rate). In the event of an SFOEWPT, this ratio should be very small signifying the fact that a slow reaction rate compared to the Universe expansion rate and a thermal nonequilibrium situation therein will prevent the reaction process of baryon-antibaryon asymmetry creation  to go in the reverse direction and thus will check the washout of the generated asymmetry.\par
\setlength{\parindent}{.5 in}   In the case of SM, the value of R is quite large \cite{47}, implying that in this model, a washout cannot be checked and thus the observed baryon-antibaryon asymmetry cannot be explained. On the other hand, in models where an SFOEWPT is possible, we may expect to have a very small value of R. As an example, in Fig.4 we show the result of calculation of R in the LHT model \cite{51}. The vast difference in the value of R in the asymmetric phase in the cases of SM($\sim$ $10^{12}$)\cite{47} and LHT ($\sim$ $10^{-21}$)\cite{51} can be understood by looking at the gauge-Higgs sectors in the two cases. In the case of LHT, we have two distinct gauge boson spectra: the heavy particles, $W^a_H$ , $B_H$ which get mass by explicit symmetry breaking and the light particles, $W^a_L$ , $B_L$ which get mass by SSB. The light gauge bosons behave like the SM gauge bosons. Since the sphaleron transition rates as well as the values of R are exponentially suppressed by the sphaleron energies which, in turn, depend on the energy functionals of the W bosons, the presence of additional heavy gauge bosons in the LHT model proves to be quite effective in making R very small. On the other hand, just after the inverse EWPT, huge baryon number violation takes place within the symmetric phase bubbles, where the rate of this violation is $10^{31}$ times higher for the T-even particles than for the T-odd ones. The details of the mechanism, being discussed here, can be found in Ref.\cite{51}.
\begin{figure}[h]
  \centering
  \includegraphics{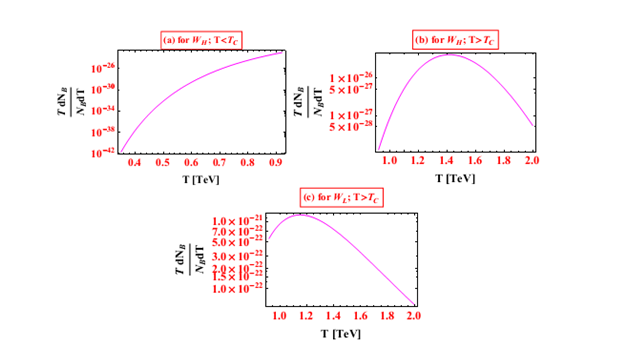}
  \caption{Ratio of the baryon number violation rate to the Universe expansion rate, (a) in the symmetric phase, (b) in the asymmetric phase for T-odd gauge boson, (c) in the asymmetric phase for T-even gauge boson.(Ref.\cite{51})(The figure is reprinted with permission from, 'On Electroweak Baryogenesis in the Littlest Higgs model with T parity', Sahazada Aziz and Buddhadeb Ghosh, \textit{Modern Physics Letters},Vol.27, No.34, Copyright@2012, World Scientific Publishing Company).}\label{Figure 4}
\end{figure}
A successful EWBG is characterized by the so-called \emph{sphaleron decoupling condition} which is a bound on the quantity, $\phi(T_c)/T_c$, $T_c$ being the transition temperature and $\phi(T_c)$ the value of the Higgs field at the transition temperature. For $m_H=126  GeV$, this bound has been derived [52] to be $\phi(T_c)/T_c>1.16$.\par
\setlength{\parindent}{.5 in} In the scenario of the inverse SFOEWPT in the LHT model, a two-step baryogenesis has been proposed \cite{51, 59}, which is schematically shown in Fig.5.
\begin{figure}[h]
  \centering
  \includegraphics{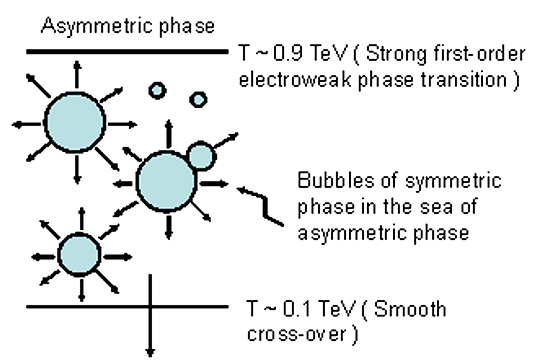}
  \caption{Electroweak bubbles in the inverse phase transition scenario in the littlest Higgs model with T parity. Inside the bubbles, the vacuum expectation value of the physical  Higgs field,$<h>=0$. Outside the bubbles,$<h>=1.1  TeV$.(Ref.\cite{59})(Reprinted with permission from [ Sahazada Aziz and Buddhadeb Ghosh, \textit{Physical Review} \textbf{D89}, 013004 (2014)] Copyright (2014) by the American Physical Society).}\label{Figure 5}
\end{figure}
In the first step an inverse SFOEWPT occurs at $T\sim0.9 TeV (t\sim10^{-13} s)$ and bubbles of symmetric phase start forming in the background of asymmetric phase. Baryon number violation takes place within these expanding  bubbles by massless T-even gauge bosons. Then at $T\sim0.1 TeV (t\sim10^{-11} s)$ there is a smooth cross-over for the T-even particles from a massless condition to a massive condition and thus baryon number violation is suppressed from this point onward.\par
\setlength{\parindent}{.5 in}  Although we are primarily considering in this article a baryogenesis scenario, which is driven by a thermally non-equilibrium EWPT, we must mention here that there are baryogenesis models that are not directly depended on EWPT and the sphaleron transitions; viz., the post-sphaleron baryogenesis\cite{53} and the TeV scale baryogenesis \cite{54}. In the former, baryon number and CP violations take place, after EWPT, in the decay of a scalar singlet into $6q$ and $6\bar q$ states. The latter is an extension of MSSM containing two new superfields: a gauge-singlet field N and a colour-triplet field X. The decay of N mediated by the exchange of X generates additional baryon asymmetry at the TeV scale. We may note that a recent theoretical study \cite{55} based on the CMS $eejj$ data\cite{56} points to the necessity of implementation of a baryogenesis mechanism below the electroweak scale.
\section{Electroweak bubbles and gravitational waves}
During the electroweak phase transition, bubbles of  ‘lower temperature phase’ are formed at the background of the ‘higher temperature’. Aspects of generation of gravitational waves (GW) from the collisions of these bubbles as well from the turbulence of the plasma have been studied in recent times \cite{57}. The intensities and frequencies of the gravitational waves will depend on the dynamics of the bubbles which in turn will be dependent on the model used for calculation. An SFOEWPT is conducive for the processes generating the GW.\par
\setlength{\parindent}{.5 in}The main input for studying the bubble dynamics is the FTEP. The pressure, being the negative of FTEP \cite{58}, can be determined both inside and outside the bubble. Expansion of the bubble will be caused by a greater inside pressure than outside. Fig.6 shows a plot of pressure difference against temperature, calculated \cite{59} in the LHT model.
\begin{figure}[h]
  \centering
  \includegraphics{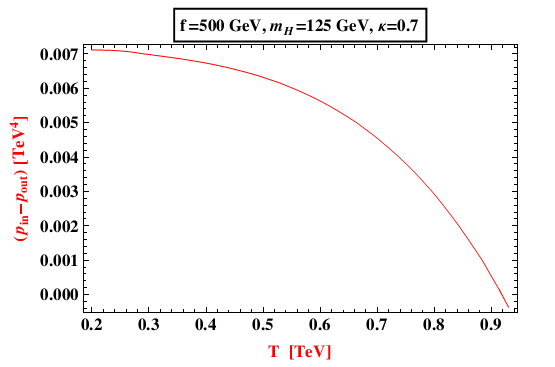}
  \caption{Excess pressure within the bubbles of the symmetric phase as a function of temperature, in the LHT model. Here $f$ is the scale and $\kappa$ is the mirror fermion coupling constant in the LHT model. (Ref.\cite{59})(Reprinted with permission from [ Sahazada Aziz and Buddhadeb Ghosh, \textit{Physical Review} \textbf{D89}, 013004 (2014)] Copyright (2014) by the American Physical Society).}\label{Figure 6}
\end{figure}
Also, the entropy, energy and enthalpy densities may be determined in terms of the pressure. Then, the velocities of the plasma inside and outside the bubble can be determined from the pressure and energy density using hydrodynamic equations \cite{60}.\par
\setlength{\parindent}{.5 in}The motion of the bubbles can be broadly classified as \emph{deflagration} or \emph{detonation} depending on whether the velocity of the plasma inside the bubble is greater or less than that outside, respectively. Also, the motion will be \emph{supersonic, Jouguet} or \emph{subsonic} if the velocity inside is greater than, equal to or less than the sound velocity in the medium, respectively.\par
\setlength{\parindent}{.5 in}Differences in the structure of FTEP in different models may yield different types of bubble motions. In the LHT model, a supersonic deflagration motion has been obtained \cite{59}.\par
\setlength{\parindent}{.5 in}The frequency of GW generated by the bubble collisions in the early Universe may be assumed to be the inverse of the time-scale of phase transition or the bubble nucleation rate, the latter being related to the excess free energy of the bubbles.
\begin{figure}[h]
  \centering
  \includegraphics{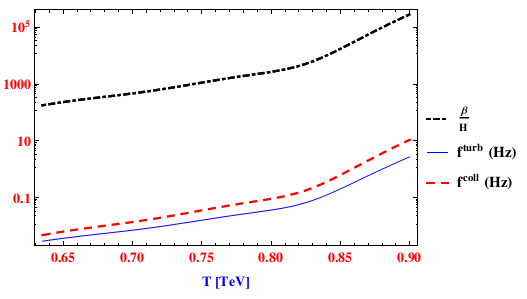}
  \caption{Plot of frequency of GW in the present Universe due to bubble collisions (dashed line) and from turbulence (solid line) for a fixed bubble wall velocity, $v_W=0.7$, for various temperatures in the early Universe.  The frequency varies as $\beta/H$ as shown by the dash-dotted line,  where $\beta$ is the inverse of time-scale of phase transition and  H is the Hubble parameter.(Ref.\cite{59})(Reprinted with permission from [ Sahazada Aziz and Buddhadeb Ghosh, \textit{Physical Review} \textbf{D89}, 013004 (2014)] Copyright (2014) by the American Physical Society).}\label{Figure 7}
\end{figure}
The frequency, thus obtained may be related to the frequency of GW in the present Universe with the help of the scale factors in the two epochs. In similar way, the intensity of GW in the present Universe can be obtained from that of the early Universe, the latter being related to the bubble wall velocity.\par
\setlength{\parindent}{.5 in}The frequency of GW generated by the stirring of the plasma or by the turbulent bulk motion of the plasma is the inverse of the so-called ‘stirring scale’[61], which is again related to the bubble radius, bubble wall velocity and the bubble nucleation time scale. The details of the derivations of the expressions of GW frequency and intensity and related analyses can be found in Ref.\cite{59}. In Fig. 7 and 8, we have shown plots of frequencies and intensities of GW calculated in the framework of the LHT model. The GW intensities appear to be quite small and the frequencies are in the deci-hertz range. Such small intensities and frequencies are in the range of future GW detectors, such as the Ulimate Deci-Hertz Interferometer Gravitational Wave Observatory or the Big Bang Observer Correlated \cite{61}.
\newpage
\begin{figure}[h]
  \centering
  \includegraphics{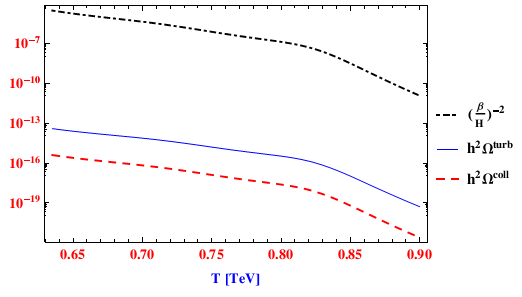}
  \caption{Plot of intensity of  GW in the present Universe due to bubble collisions (dashed line) and from turbulence (solid line) for a fixed bubble wall velocity, $v_W=0.7$, for various temperatures in the early Universe.  The intensity $(\omega=\frac{\rho_{GW}}{\rho_{tot}}),\rho$ being the energy density, varies as $(\beta/H)^{-2}$ as shown by the dash-dotted line. (Ref.\cite{59})(Reprinted with permission from [ Sahazada Aziz and Buddhadeb Ghosh, \textit{Physical Review} \textbf{D89}, 013004 (2014)] Copyright (2014) by the American Physical Society).}\label{Figure 8}
\end{figure}

\section{Conclusions}
In conclusion, we have highlighted in this article,  some past and present studies of EWPT, baryogenesis, electroweak bubble dynamics and GW generation. In view of the considered inadequacy of the SM so far as the description of the cosmological phenomena are concerned, but at the same time the recent discovery of an SM-like Higgs at the LHC, we should  now be poised for surprises from the results of the forthcoming LHC runs.
\section*{Acknowledgements}
The author wishes to thank Prof. D. Choudhury for providing the impetus to write this article. He also acknowledges many useful discussions with Dr. S. Aziz.
\pagebreak


\begin{thebibliography}{55}
\bibitem{1}
E. W. Kolb and M.S. Turner,
\textit{The Early Universe, Frontiers in Physics}.
(Westview, Boulder, CO, 1994)
\bibitem{2}
CDF and D0 Collaborations,
\textit{Phys. Rev. Lett}.
\textbf{109}, 071804 (2012)  hep-ex/1207.6436.\\
The Atlas Collaboration,
\textit{Phys. Lett.}.
\textbf{B716}, 1(2012) hep-ex/1207.7214\\
The CMS Collaboration,
\textit{Phys. Lett.}
\textbf{B716}, 30(2012) hep-ex/1207.7235
\bibitem{3}
ALEPH, DELPHI, L3 and OPAL Collaborations; The LEP Working Group for Higgs Boson Searches,
\textit{Phys. Lett.}
\textbf{B565}, 61(2003)  hep-ex/0306033\\
DELPHI Collaborations,
\textit{Eur. Phys.}
\textbf{J 32}, 145(2004) hep-ex/0303013\\
\bibitem{4}
C. Grojean, G. Servant and J. D. Wells,
\textit{Phys. Rev.}
\textbf{D71}, 036001(2005) hep-ph/0407019\\
S. Profumo, M. J. Ramsay-Musolf  and G. Shaughnessy,
\textit{JHEP}
\textbf{0708}, 010(2007) hep-ph/0705.2425\\
J. M. Cline, M. Jarvinen and F. Sannino,
\textit{Phys. Rev}
\textbf{D78}, 075027(2008)  hep-ph/0808.1512\\
A. Ashoorioon and T. Constandin,
\textit{JHEP}
\textbf{0907}, 086(2009) hep-ph/0904.0353\\
M. Jarvinen, T. A. Ryttov and F. Sannino,
\textit{Phys. Rev.}
\textbf{D79}, 095008(2009)\\
J. M. Cline, G. Laporte, H.Yamashite and S. Kraml,
\textit{JHEP}
\textbf{0907}, 040(2009) hep-ph/0905.2559\\
V. Barger, D. J. H. Chung, A. J. Long and L.-T. Wang
\textit{Phys. Lett.}
\textbf{B710}, 1(2012) hep-ph/1112.5460\\
\bibitem{5}
D. J. H. Chung, A. J. Long and L-T. Wang,
\textit{Phys. Rev.}
\textbf{D87}, 023509 (2013) hep-ph/1209.1819\\
V. Q. Phong, H. N. Long, V. T. Van,
\textit{Phys. Rev.}
\textbf{D88}, 096009(2013) hep-ph/1309.0355\\
A. Katz and M. Perelstein,
\textit{JHEP}
\textbf{07}, 108(2007) hep-ph/1401.1827\\
S. Profumo, M. J. Ramsay-Musolf , C. L. Wainwright and P. Winslow,
\textit{Phys. Rev}
\textbf{D91}, 035018(2015) hep-ph/1407.5342\\
A. Farzinnia and J. Ren,
\textit{Phys. Rev.}
\textbf{D90}, 075012(2014) hep-ph/1408.3533\\
F. Sannino and J. Virkaj\"{a}rvi,
\textit{Phys. Rev.}
\textbf{D92}, 045015(2015) hep-ph/1505.05872
\bibitem{6}
D. A. Kirzhnitz and A. D. Linde,
\textit{Annals of Physics}
\textbf{101}, 195(1976)
\bibitem{7}
J. Iliopoulos and N. Papanicolaou
\textit{Nucl. Phys.}
\textbf{B111}, 209(1976)
\bibitem{8}
A. H. Guth and E. J. Weinberg,
\textit{Phys. Rev. Lett.}
\textbf{45}, 1131 (1980)
\bibitem{9}
S. Coleman and E. Weinberg,
\textit{Phys. Rev.}
\textbf{D7}, 1888(1973)
\bibitem{10}
E. Witten,
\textit{Nucl. Phys.}
\textbf{B177},477(1981)
\bibitem{11}
M. A. Sher,
\textit{Phys. Rev.}
\textbf{D22}, 2989(1980)
\bibitem{12}
D. Bailin and A. Love,
\textit{Introduction to Gauge Field Theory}
(Adam and Hilger, Bristol and Boston, 1986)
\bibitem{13}
A. Zee,
\textit{Quantum Field Theory in a Nutshell}
(Princeton University Press, USA, 2003)
\bibitem{14}
P. Arnold and O. Espinosa,
\textit{Phys. Rev.}
\textbf{D47}, 3546(1993) hep-ph/9212235\\
J. E. Bagnasco and M. Dine,
\textit{Phys. Lett.}
\textbf{B303}, 308(1993)
\bibitem{15}
B. Bunk, E. M. Ilgenfritz,  J. Kripfgang and A. Schiller,
\textit{Phys. Lett.}
\textbf{B284}, 371(1992);
\textit{Nucl. Phys.}
\textbf{B403}, 453(1993)\\
E. –M. Ilgenfritz,  J. Kripfgang, H. Perlt and A. Schiller,
\textit{Phys. Lett.}
\textbf{B356}, 561(1995)  hep-lat/9506023\\
M. G\"{u}rtler, E. –M. Ilgenfritz,  J. Kripfgang, H. Perlt and A. Schiller,
\textit{Nucl. Phys.}
\textbf{B483},383(1997) hep-lat/9605042
\bibitem{16}
K. Kajantie, K. Rummukainen and M. Shaposnikov,
\textit{Nucl. Phys.}
\textbf{B407}, 356(1993) hep-ph/9305345
\bibitem{17}
K. Farakos, K. Kajantie, K. Rummukainen and M. Shaposnikov,
\textit{Phys. Lett.}
\textbf{B336}, 494(1994)  hep-ph/9405234
\bibitem{18}
Z. Fodor, J. Hein, K. Jansen, A. Jaster and I. Montvay,
\textit{Nucl. Phys.}
\textbf{B439}, 147(1995) hep-lat/9409017
\bibitem{19}
K. Kajantie, M. Laine, K. Rummukainen and M. Shaposnikov,
\textit{Phys. Rev. Lett.}
\textbf{77}, 2887 (1996) ;
\textit{Nucl. Phys.}
\textbf{B493}, 413(1997) hep-lat/9612006
\bibitem{20}
F. Csikor and Z. Fodor,
\textit{Nucl. Phys.}
\textbf{B474}, 421(1996) hep-lat/9601016
\bibitem{21}
W. Buchmüller, Z. Fodor and A. Hebecker,
\textit{Nucl. Phys.}
\textbf{B447}, 317(1995) hep-ph/9502321
\bibitem{22}
R. Rangarajan, S. Sengupta and A. M. Srivastava,
\textit{Astropart. Phys.}
\textbf{17}, 167 (2002)  hep-ph/9911488
\bibitem{23}
R. Brandenberger, A. –C. Davis and M. Trodden,
\textit{Phys. Lett.}
\textbf{B335}, 123(1994)  hep-ph/9403215 \\
B. Layek, S. Sanyal and A. M. Srivastava,
\textit{Int. J. Mod. Phys.}
\textbf{A18}, 4851(2003) hep-ph/0107174
\bibitem{24}
T. Prokopec, R. Brandenberger and  A. –C. Davis,
hep-ph/9601327
\bibitem{25}
See, for example, S. P. Martin, hep-ph/9709356
\bibitem{26}
J. R. Espinosa, M. Quir\'os and F. Zwirner,
\textit{Phys. Lett.}
\textbf{B307}, 106(1993)  hep-ph/9303317\\
A. Bignole, J. R. Espinosa, M. Quir\'os and F. Zwirner,
\textit{Phys. Lett}
\textbf{B324}, 181(1994)  hep-ph/9312296\\
J. M. Cline and G. D. Moore,
\textit{Phys. Rev. Lett.}
\textbf{81}, 3315 (1998)  hep-ph/9806354\\
W. Huang, J. Shu and Y. Zhang,
\textit{JHEP}
\textbf{1303}, 164(2013) hep-ph/1210.0906
\bibitem{27}
J. M. Cline and K. Kainulainen,
\textit{Nucl. Phys.}
\textbf{B482}, 73(1996) hep-ph/9605235; \emph{ibid}, \textbf{510}, 88(1998)\\
M. Laine and K. Rumukainen,
\textit{Phys. Rev. Lett.}
\textbf{80}, 5259 (1998)  hep-ph/9804255\\
F. Csikor, Z. Fodor, P. Hegedus, A. Zakovac, S. D. Katz and A. Piroth
\textit{Phys. Rev. Lett}
\textbf{85}, 923 (2000)  hep-ph/0001087
\bibitem{28}
S. W. Ham, S. K. Oh and D. Son,
\textit{Phys. Rev.}
\textbf{D71}, 015001(2005) hep-ph/0411012\\
R. Fok and G. D. Kribs,
\textit{Phys. Rev.}
\textbf{D78}, 075023(2008) hep-ph/0803.4207
\bibitem{29}
S. W. Ham and S. K. Oh,
\textit{Phys. Rev.}
\textbf{D76}, 095018(2007) hep-ph/0708.1785
A. Alriche and S. Nasri,
\textit{Phys. Rev.}
\textbf{D83}, 045032(2011) hep-ph/1008.3106
\bibitem{30}
D. J. H. Chung and A. J. Long,
\textit{Phys. Rev.}
\textbf{D81}, 123531(2010) hep-ph/1004.0942
\bibitem{31}
J. Kozaczuk, S. Profumo and C. L. Wainwright,
\textit{Phys. Rev.}
\textbf{D87}, 075011(2013) hep-ph/1302.4781\\
W. Huang, Z. Kang, J. Shu, P. Wu and J. M. Yang,
\textit{Phys. Rev.}
\textbf{D91}, 025006(2015) hep-ph/1405.1152\\
J. Kozaczuk, S. Profumo, L. S. Haskins and C. L. Wainwright,
\textit{JHEP}
\textbf{01}, 144(2015) hep-ph/1407.4134\\
X-J Bi, L. Bian, W. Huang, J. Shu and P-F Yin,
\textit{Phys. Rev.}
\textbf{D92}, 023507(2015) hep-ph/1503.03749\\
\bibitem{32}
M. Carena, N. R. Sha and C. E. M. Wagner
\textit{Phys. Rev.}
\textbf{D85}, 036003(2012) hep-ph/1110.4378\\
J. M. Cline and K. Kainulainen,
\textit{Phys. Rev.}
\textbf{D87}, 071701(2013) hep-ph/1302.2614\\
H. H. Patel and M. J. Ramsay-Mousolf,
\textit{Phys. Rev.}
\textbf{D88}, 035013(2013) hep-ph/1212.5652\\
A. Alriche and S. Nasri,
\textit{JCAP}
\textbf{07}, 035(2013) hep-ph/1304.2055\\
M. Fairburn and R. Hogan,
\textit{JHEP}
\textbf{09}, 022(2013) hep-ph/1305.3452\\
M. Jiang, L. Bian, W. Huang and J. Shu, hep-ph/1502.07574\\
\bibitem{33}
J. M. Cline and P. -A. Lemieux,
\textit{Phys. Rev.}
\textbf{D55}, 3873(1997) hep-ph/960920\\
G. C. Dorch, S. J. Huber and J. M. No,
\textit{JHEP}
\textbf{10}, 029(2013) hep-ph/1305.6610\\
G. C. Dorch, S. J. Huber, K. Mimasu and J. M. No,
\textit{Phys. Rev. Lett.}
\textbf{113}, 211802(2014) hep-ph/1405.5537\\
\bibitem{34}
G. Panico and M. Serone,
\textit{JHEP}
\textbf{05}, 024(2005) hep-ph/0502255\\
N. Maru and K. Takenaga,
\textit{Phys. Rev.}
\textbf{D72}, 04603(2005) hep-ph/0505066
\bibitem{35}
M. Sakamoto and K. Takenaga,
\textit{Phys. Rev.}
\textbf{D80}, 085016(2009) hep-ph/0908.0987
\bibitem{36}
S. Aziz, B. Ghosh and G. Dey,
\textit{Phys. Rev.}
\textbf{D79}, 075001(2009) hep-ph/0901.3442
\bibitem{37}
See, for example, M. Perelstein,
\textit{Prog. Part. Nucl. Phys.}
\textbf{D58}, 247(2007) hep-ph/0512128
\bibitem{38}
A. Noble and M. Perelstein,
\textit{Phys. Rev.}
\textbf{D78}, 063518(2008) hep-ph/0711.3018
\bibitem{39}
J. Beringer et al. [Particle Data Group Collaboration],
\textit{Phys. Rev.}
\textbf{D86}, 010001(2012)
\bibitem{40}
M. Shaposnikov,
\textit{J. of Phys. Conference Series}
\textbf{171}, 012005(2009)
\bibitem{41}
H. Toki and H. Kawai,
\textit{Prog. Theor. Phys.}
\textbf{98}, 449(1997) hep-ph/9703421\\
H. Davoudiasl, R. Kitano, G. D. Krib, H. Muryama and P. J. Steinhardt,
\textit{Phys. Rev. Lett.}
\textbf{93}, 201301 (2004)  hep-ph/0403019\\
G. L. Alberghi,
\textit{Phys. Rev.}
\textbf{D81}, 10351(2010) hep-ph/1002.3713\\
S. K. Modak and D. Singleton,
\textit{Eur. Phys. J.}
\textbf{C75}, 5(2015) gr-qc/1410.6785\\
\bibitem{42}
E. W. Kolb, A. Riotto and I. I. Tkachev,
\textit{Phys. Lett.}
\textbf{B423}, 348(1998)  hep-ph/9801306\\
J. A. Lopez-Perez and N. Rius, hep-ph/0404124\\
M. Trodden, hep-ph/0411301\\
\bibitem{43}
M. Trodden,
\textit{Rev. Mod. Phys.}
\textbf{71}, 1463(1999)  hep-ph/9803479\\
J. M. Cline and K. Kainulainen,
\textit{Phys. Rev. Lett.}
\textbf{85}, 5519 (2000)  hep-ph/00002272\\
J. M. Cline, M. Joice and K. Kainulainen,
\textit{JHEP}
\textbf{07}, 018(2000) hep-ph/0006119\\
M. Losada,
\textit{Nucl. Phys.}
\textbf{B537}, 3(1999) hep-ph/9806519\\
M. Quir\'os,
\textit{J. Phys. A: Math Theo.}
\textbf{40}, 6573(2007)\\
M. Carena, G. Nardini, M. Quir\'os and C. E. M. Wagner,
\textit{Nucl. Phys.}
\textbf{B812}, 243(2009) hep-ph/0809.3760\\
K. Funakubo and E. Senaha,
\textit{Phys. Rev.}
\textbf{D79}, 115024(2009) hep-ph/0905.2022\\
V. Cirigliano, Y. Li, S. Profumo and M. H. Ramsay-Mousolf,
\textit{JHEP}
\textbf{01}, 002(2010) hep-ph/0910.4589
\bibitem{44}
J. Kang, P. Langacker, T. Li and T. Liu,
\textit{JHEP}
\textbf{04}, 097(2011) hep-ph/0911.2939\\
S. W. Ham, S. -A. Shin and S. K. Oh, hep-ph/1001.1129\\
K. Blum, C. Delaunay, M. Losada, Y. Nir and S. Tulin,
\textit{JHEP}
\textbf{05}, 101(2010) hep-ph/1003.2447
\bibitem{45}
G. Grant and M. Hindmarch,
\textit{Phys. Rev.}
\textbf{D64}, 016002(2001) hep-ph/0101120\\
S. Kanemura, Y. Okada and E. Senaha,
\textit{Phys. Lett}
\textbf{B606}, 361(2005)  hep-ph/0411354\\
A. Tranberg and B. Wu,
\textit{JHEP}
\textbf{07}, 087(2012) hep-ph/1203.5012\\
K. Enqvist, P. Stephens, O. Taanilla and A. Tranberg,
\textit{JCAP}
\textbf{1009}, 019(2010) astro-phy.CO/1005.0752\\
J.M.No, hep-ph/1504.07840
\bibitem{46}
A. D. Sakharov,
\textit{JETP Lett}
\textbf{91B}, 24(1967)
\bibitem{47}
P. Arnold and L. McLerran,
\textit{Phys. Rev.}
\textbf{D36}, 581(1987)
\bibitem{48}
F. R. Klinkhamer and N. S. Manton,
\textit{Phys. Rev.}
\textbf{D30}, 2212(1984)
\bibitem{49}
M. Quir\'os, hep-ph/9901312
\bibitem{50}
J. M. Cline, hep-ph/0609145
\bibitem{51}
S. Aziz and B. Ghosh,
\textit{Mod. Phys. Lett.}
\textbf{A27}, 1250190 (2012), hep-ph/1007.0485
\bibitem{52}
E. Senaha,
\textit{Toyama International Workshop on Higgs as a Probe of New Physics},
2013, 13-16 February,2013 hep-ph/1305.1563
\bibitem{53}
K. S. Babu, R. N. Mohapatra and S. Nasri,
\textit{Phys. Rev. Lett}
\textbf{97}, 131301(2006)  hep-ph/0606144\\
K. S. Babu, P. S. Bhupal Dev and R. N. Mohapatra,
\textit{Phys. Rev.}
\textbf{D79}, 015017(2009)hep-ph/0811.3411\\
K. S. Babu, P. S. Bhupal Dev, Elaine C. F. S. Fortes and R. N. Mohapatra,
\textit{Phys. Rev.}
\textbf{D87}, 115019(2013) hep-ph/1303.6918
\bibitem{54}
K. S. Babu, R. N. Mohapatra and S. Nasri,
\textit{Phys. Rev. Lett}
\textbf{98}, 161301(2007)  hep-ph/0612357
\bibitem{55}
M. Dhuria, C. Hati, R. Rangarajan and U. Sarkar,
\textit{Phys. Rev.}
\textbf{D91}, 055010(2015) hep-ph/1501.04815
\bibitem{56}
CMS Collaboration,
\textit{CMS-PAS-EXO}-12-041
\bibitem{57}
S. J. Huber and T. Constandin,
\textit{JCAP}
\textbf{09}, 022(2008) hep-ph/0806.1828\\
C. Caprini, R. Durer and G. Servant,
\textit{Phys. Rev.}
\textbf{D77}, 124015(2008) hep-ph/0711.2593\\
A. M\'egevand,
\textit{Phys. Rev.}
\textbf{D78}, 084003(2008) astro-ph.CO/0804.0391\\
T. Kahniasvili, A. A. Kosowski, G. Gogoberidze and Y. Macavin,
\textit{Phys. Rev.}
\textbf{D78}, 043003(2008) astro-ph/0806.0293\\
T. Kahniasvili, L. Kisslinger and T. Stevens,
\textit{Phys. Rev.}
\textbf{D81}, 023004(2010) astro-ph.CO/0905.0643\\
J. M. No,
\textit{Phys. Rev.}
\textbf{D84}, 124025(2011) hep-ph/1103.2159\\
H. L. Child, J. T. Giblin Jr.,
\textit{JCAP}
\textbf{10}, 001(2010) astro-ph.CO/1207.6408\\
N. Kitajima and F. Takahashi,
\textit{Phys. Lett}
\textbf{B745}, 112(2015)  hep-ph/1502.03725\\
A. Kamada and M. Yamada, hep-ph/1505.01167\\
J. Kozaczuk, hep-ph/1506.04741
\bibitem{58}
M. E. Carrington and J. L. Kapusta,
\textit{Phys. Rev.}
\textbf{D47}, 5304(1993)
\bibitem{59}
S. Aziz and B. Ghosh,
\textit{Phys. Rev.}
\textbf{D89}, 013004(2014) hep-ph/1304.2997
\bibitem{60}
J. R. Espinosa, T. Constandin, J. M. No and G. Servant,
\textit{JCAP}
\textbf{06}, 028(2010) hep-ph/1004.4187\\
L. D. Landau and E. M. Lifshitz,
\textit{Fluid Mechanics},
(Pergamon Press, New York, 1989)\\
A. M\'egevand and A. D. Sanchez,
\textit{Nucl. Phys.}
\textbf{B820}, 47(2009) hep-ph/0904.1753
\bibitem{61}
L. Leitao,A. M\'egevand and A. D. Sanchez,
\textit{JCAP}
\textbf{10}, 024(2012) astro-ph.CO/1205.3070

\end{thebibliography}
\end{document}